\documentclass{aa}  
\usepackage{graphicx}
\usepackage{txfonts}
\usepackage{natbib}
\bibpunct{(}{)}{;}{a}{}{,}
\begin{document}
   \title{Discovery of a new cataclysmic variable through optical variability and X-ray emission}

   \subtitle{}

   \author{Magaretha L. Pretorius
          \and
          Christian Knigge
          }

   \offprints{M.L. Pretorius}

   \institute{School of Physics and Astronomy, University of Southampton, Highfield, Southampton SO17 1BJ, United Kingdom\\
              \email{mlp@astro.soton.ac.uk, christian@astro.soton.ac.uk}
             }
   \authorrunning{M.L. Pretorius \& C. Knigge}


 
  \abstract
   {}
   {We present discovery observations of the new cataclysmic variable star (CV) 1RXS J092737.4-191529, as well as spectra and photometry of SY Vol.  The selection technique that turned up these two CVs is described; it should be efficient for finding dwarf novae with high outburst duty cycles.
}
   {Two very common observational features of CVs, namely optical variability and X-ray emission, are combined to select targets for follow-up observations.  Long-slit spectra were taken to identify CVs in the sample.
}
   {Two out of three objects selected in this way are CVs.  One of these is the known dwarf nova SY Vol, while the second system, 1RXS J092737.4-191529, is a new discovery.  We present medium resolution spectra, $UBVRI$ magnitudes, and high-speed photometry for both these CVs.  Rapid flickering in the light curve of 1RXS J092737.4-191529 confirms the mass transferring binary nature of this object; it is probably a dwarf nova that was in quiescence during our observations.
}
   {}

   \keywords{Stars: dwarf novae -- novae, cataclysmic variables}

   \maketitle

\section{Introduction}
Cataclysmic variable stars (CVs) are semi-detached binary stars, consisting of a white dwarf primary accreting from a companion that is usually a late-type, approximately main-sequence star.  The orbital periods ($P_{orb}$) of these binaries are typically on the order of hours.  \citet{bible} gives a comprehensive review of the subject.

Based on their long term photometric behaviour, non-magnetic CVs are divided into the classes dwarf novae (DNe) and nova-like variables (NLs).  The physical distinction between them is that the time-averaged mass transfer rate ($\dot{M}$) is lower in DNe than in NLs.  The majority of CVs are DNe; the defining characteristic of this class is outbursts during which the system brightens by typically 2--5 mag.  Outburst durations are days to tens of days, and the recurrence interval ranges from tens of days to decades.  NLs are usually in a state of high $\dot{M}$, but a subtype of NLs, called VY Scl stars, are characterised by occasional low states (at least 1~mag fainter than the average brightness, lasting for weeks to years; see \citealt{WadeWard85}).  DN outburst are caused by an accretion disc instability (e.g. \citealt{Osaki96}), while VY Scl star low states probably result from a reduction in the rate at which the secondary star loses mass (e.g. \citealt{LivioPringle94}).

Many magnetic CVs (see \citealt{Cropper90} and \citealt{Patterson94} for reviews) vary by several magnitudes.  In the case of polars, this is probably caused by the same mechanism as in VY Scl stars, while the high states seen in some intermediate polars (IPs) are (in at least some cases) believed to be caused by the disc instability that operates in DNe (\citealt{KimWheelerMineshige92}; \citealt{brian96}; \citealt{HellierMukaiBeardmore97}).

In addition to the mechanisms mentioned above, large amplitude variability in CVs is caused by nova eruptions, which occur in all types of hydrogen-rich CVs.  All CVs are therefore expected to be large amplitude variables, although the recurrence times and duty cycles of the variability ranges widely.

A second attribute that should be shared by all active CVs is X-ray emission generated in the accretion flow.  X-ray observations of CVs indicate $10^{29}\,\mathrm{erg\,s^{-1}} \la  L_x \la 3 \times 10^{32}\,\mathrm{erg\,s^{-1}}$ (e.g. \citealt{PattersonRaymond85}; \citealt{vanTeeselingBeuermannVerbunt96}; \citealt{VerbuntBunkRitter97}; \citealt{WheatleyBurleighWatson00}).

Given that large amplitude variability and X-ray emission are ubiquitous features of CVs, a combination of these two criteria is a promising way to identify CVs.  Wide field photographic sky surveys have recorded large amplitude variability in many stars; the nature of the vast majority of these variables is not known.  Furthermore, optical identifications exist for only a small fraction of all sources detected in ROSAT All-Sky Survey (the RASS; see \citealt{rosatbsc} and \citealt{rosatfsc}).  There should be many CVs amongst the unidentified sources in this survey.  More than 100 CVs have already been discovered by ROSAT \citep{Gansicke05}.  Optical follow-up of a sample of bright ($>0.2\,\mathrm{counts/s}$) ROSAT sources led to the discovery of 11 CVs \citep{SchwopeBrunnerBuckley02}.  In a search aimed specifically at CVs, \cite{GansickeMarshEdge05} use a fainter ROSAT flux limit together with the requirements $J-K<1$, $K>11$, and $R<17$.  

Here, we investigate the feasibility of selecting CVs for large amplitude optical variability and X-ray emission.  We present our selection procedure and observations in the next two sections, before discussing and summarising the results in Section~\ref{sec:discussion}.

\begin{table*}
 \centering
 \begin{minipage}{145mm}
 \caption[]{Log of the spectroscopic observations.  The fourth column gives the photographic $R$ magnitude of the fainter of the two epochs, as well as the difference between the two $R$-band measurements.  The dates are for the start of the night; coordinates are for the optical variables.}
  \label{tab:logsp}
  \begin{tabular}{lllllll}
  \hline
  \noalign{\smallskip}
Object      & $\alpha_{2000}$ & $\delta_{2000}$ & $R$ $(\Delta R)$ & Date & HJD $2453000.0+$ & $t_{int}$/s \\
  \noalign{\smallskip}
  \hline
  \noalign{\smallskip}
1RXS J085325.4-711255$^a$ & 08:53:26.2 & -71:12:48 & 17.0 (2.4) &  5 Jan 2004 & 376.47567 & 1400  \\
                          &            &           &            & 10 Jan 2004 & 381.50743 & 1800  \\
1RXS J092737.4-191529     & 09:27:37.1 & -19:15:34 & 18.3 (2.9) &  5 Jan 2004 & 376.52463 & 1600  \\
                          &            &           &            & 10 Jan 2004 & 381.57730 & 2000  \\
1RXS J152912.9-101623     & 15:29:12.2 & -10:16:28 & 18.4 (2.0) &  4 Mar 2006 & 799.55906 & 1300  \\
  \noalign{\smallskip}
  \hline
  \end{tabular}
Notes: $^a$This is SY Vol; $t_{int}$ is the integration time. \hfill
 \end{minipage}
\end{table*}

\section{Selection}
Our selection is aimed at X-ray sources that can be associated with optical variables.  We use two epochs of photographic images to identify large amplitude optical variables, and correlate these objects with sources detected in the RASS.  Since we use only two observations to search for variability, we expect to find mainly objects with frequent, high duty cycle brightness variations.

The photographic plate digitising machine, SuperCOSMOS, has scanned Schmidt sky survey plates taken with the UK, European Southern Observatory (ESO), and Palomar Schmidt telescopes.  The resulting data are publicly available and comprise the SuperCOSMOS Sky Survey (the SSS; see \citealt{supercosmosI}; \citealt{supercosmosII}; \citealt{supercosmosIII}).  The survey covers the southern celestial hemisphere ($\delta < +3.0$) at two epochs in $R$, as well as at a single epoch in $B$ and $I$.  The data can be accessed most readily through the SuperCOSMOS Science Archive \citep{HamblyReadMann04}.

Considering only a subset\footnote{This is the so-called ReliableStars subset, which consists of sources with small ellipticity, detected on at least 3 out of the 4 plates.  Objects that were assigned quality flags indicating a possibly bad image during the image analysis are excluded, together with objects that were deblended.  The ReliableStars are further restricted to areas away from the galactic plane and galactic bulge ($|b|>10^\circ$ and $>20^\circ$ from the Galactic Centre), to avoid severely crowded fields.} of sources detected in the SSS that consists mainly of genuine stars (rather than extended objects or plate defects) we choose objects that varied by more than 2 mag between the two $R$-band observations.  In addition, we require $R<19$ in both epochs.  This means that it should be possible to obtain follow-up observations with a 2-m class telescope, and excludes stars with high proper motion (without this requirement, many high proper motion objects are included, since they are detected in only one epoch).

A total of 2\,757 targets results from this selection.  We inspected scans of both $R$-band plates for 100 of these objects---in only two out of these 100 cases is the variability clearly spurious, with obvious defects on the plates very near the stellar images.  However, we have found that even apparent differences of $\ga 2$~mag are occasionally caused by poor calibration between the different plates.  Only 36 of the 2\,757 objects selected in this way can be matched to known variable stars in the SIMBAD Astronomical Database.  This underlines the fact that a large number of stars that show large amplitude variability remain completely unstudied.  Amongst the known variable stars in this sample are 8 CVs\footnote{They are WX Hyi, BI Ori, SY Vol, TU Ind, GS Pav, V803 Cen, SDSS J040714.78-064425.2, and SDSS J161332.56-000331.0.  The two systems WX Hyi and V803 Cen are known to be the optical counterparts of RASS sources \citep{VerbuntBunkRitter97}.}, 13 Mira variables, and 3 young stellar objects.  Also included in the sample of variables are two quasars; both were detected by ROSAT.

Three of the optical variables found in the SSS are close positional matches to ROSAT sources that have not been optically identified; this includes the known CV SY Vol, which is very likely the optical counterpart of 1RXS J085325.4-711255.  These three objects are selected for the present study.

\section{Observations}
\begin{figure}
 \centering
 \includegraphics[width=6.5cm]{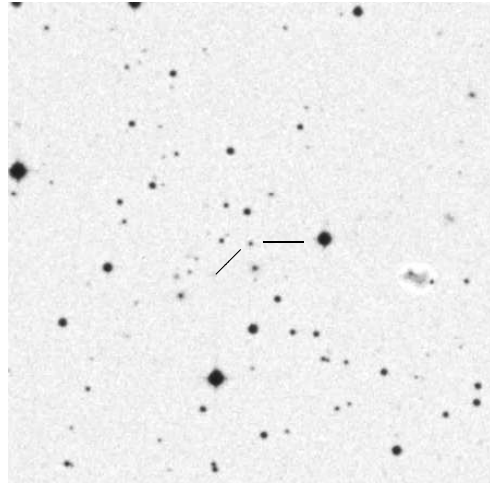}
 \caption{A $5\arcmin \times 5\arcmin$ finding chart of the optical counterpart of 1RXS J092737.4-191529.  This was made using a digitised ESO Schmidt telescope $R$-band plate.  North is at the top and east to the left of the figure.  The system is near minimum light in this image.}
 \label{fig:find}
\end{figure}

We obtained follow-up observations at the Sutherland site of the South African Astronomical Observatory (SAAO).  In addition to SY Vol, 1RXS J092737.4-191529 proved to be a CV.  A finding chart for this newly discovered CV is displayed in Fig.~\ref{fig:find}. 

Identification spectra were taken with the Grating Spectrograph on the SAAO 1.9-m telescope.  The spectrograph is equipped with an SITe charge-coupled device (CCD) detector, and was used with grating no.\,7 and a slit width of $1\farcs8$, yielding a resolution of $\simeq 5$~\AA\ over a range of $\simeq 3\,550$~\AA.  Table~\ref{tab:logsp} gives a log of the spectroscopic observations, together with positions and photographic $R$-band magnitudes of the targets.

The variable star that we associate with 1RXS J152912.9-101623 was fainter than $R \simeq 19.0$ when we observed it, and close to the magnitude limit of the spectrograph.  The S/N in the spectrum is too low to allow for a reliable determination of the spectral type, but, since no emission lines were detected, it is unlikely that this object is a CV.  

\begin{table}
 \caption[]{$UBVRI$ magnitudes of SY Vol and 1RXS J092737.4-191529.}
  \label{tab:ubvri}
  \begin{tabular}{lll}
  \hline
  \noalign{\smallskip}
    & SY Vol            & 1RXS J092737.4-191529 \\
  \noalign{\smallskip}
  \hline
  \noalign{\smallskip}
$U$ & $17.66 \pm 0.06$  & $18.05 \pm 0.08$ \\ 
$B$ & $18.46 \pm 0.05$  & $18.96 \pm 0.06$ \\
$V$ & $18.08 \pm 0.02$  & $18.60 \pm 0.04$ \\ 
$R$ & $17.60 \pm 0.02$  & $18.00 \pm 0.03$ \\ 
$I$ & $16.88 \pm 0.01$  & $17.24 \pm 0.02$ \\ 
  \noalign{\smallskip}
  \hline
  \end{tabular}
\end{table}

\begin{table*}
 \centering
 \begin{minipage}{135mm}
 \caption[]{Log of the time-resolved photometric observations.  Dates are for the start of the night; HJD is for the middle of the first integration; the average magnitude for each run is listed in the final column.}
 \label{tab:obsph}
  \begin{tabular}{llllllll}
  \hline
  \noalign{\smallskip}
Object            &  Run no. & Date & HJD $2453000.0+$ & Length/h & $t_{int}$/s & pseudo-$V$ \\
  \noalign{\smallskip}
  \hline
  \noalign{\smallskip}
SY Vol                & RP13 & 13 Jan 2005 & 384.45133 & 1.25 & 20     & 17.7  \\
                      & RP22 & 15 Jan 2005 & 386.37182 & 5.42 & 20, 26 & 17.9  \\
                      & RP24 & 16 Jan 2005 & 387.34376 & 1.32 & 20     & 17.9  \\
                      & RP26 & 17 Jan 2005 & 388.37330 & 2.35 & 20     & 18.1  \\
1RXS J092737.4-191529 & RP16 & 13 Jan 2005 & 384.56121 & 3.40 & 24     & 18.1  \\
                      & RP20 & 14 Jan 2005 & 385.45315 & 0.63 & 30     & 18.2: \\
                      & RP25 & 16 Jan 2005 & 387.42975 & 4.37 & 20     & 18.2  \\
                      & RP27 & 17 Jan 2005 & 388.50430 & 2.28 & 20     & 18.3  \\
  \noalign{\smallskip}
  \hline
  \end{tabular}
Notes: $t_{int}$ is the integration time (the photometer is a frame transfer CCD, so that there is no dead time between exposures); `:' denotes an uncertain value.\hfill
 \end{minipage}
\end{table*}

In addition to the identification spectroscopy, we obtained $UBVRI$ photometry as well as high speed photometry of SY Vol and 1RXS J092737.4-191529, using the University of Cape Town CCD photometer (the UCT CCD; see \citealt{uctccd}) on the SAAO 1-m telescope.  The $UBVRI$ magnitudes of the CVs are listed in Table~\ref{tab:ubvri}.  Each of these measurements is the average of three exposures separated by a few minutes; Table~\ref{tab:ubvri} gives the photometric errors, but we emphasise that these are smaller than the amplitude of flickering (at least in white light).  The $U-B$ and $B-V$ colours of both systems are well within the range observed in quiescent DNe \citep{BruchEngel94}.  Table~\ref{tab:obsph} gives a log of the time-resolved photometry.  These observations were made in white light; with the UCT CCD this gives photometry with an effective wavelength similar to Johnson $V$, but with a very broad bandpass.  The non-standard flux distribution of CVs and the use of white light means that the high-speed observations cannot be precisely placed on a standard photometric system.  The magnitude calibration to `pseudo-$V$' approximates Johnson $V$ to within $\simeq 0.1$~mag; colour terms were neglected in the atmospheric extinction corrections.

\subsection{SY Vol}
SY Vol is a little studied dwarf nova ranging in brightness from $V \simeq 14.4$ to $V \simeq 18.3$ \citep{surveyV}.  \cite{CieslinskiSteinerJablonski98} show a spectrum of SY Vol taken in outburst, while high-speed photometry in quiescence as well as outburst is presented by \cite{surveyV}.  The outburst recurrence timescale and orbital period of the system are not known.

\begin{figure}
 \centering
 \includegraphics[width=8.8cm]{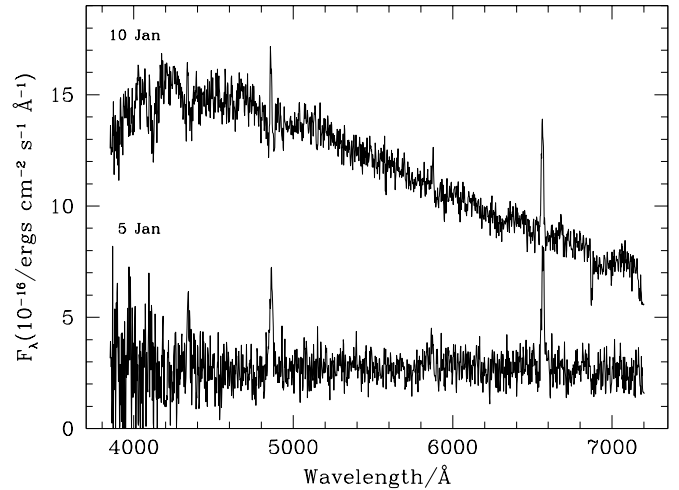}
 \caption{Spectra of SY Vol.  The first, taken on 5 January 2005, shows the system in quiescence.  SY Vol was probably on the decay from outburst on 10 January 2005, when the second spectrum was taken.}
 \label{fig:syvolspec}
\end{figure}

Fig.~\ref{fig:syvolspec} displays our spectra of SY Vol.  The first was taken when the system was in quiescence; it shows Balmer emission lines on a flat continuum.  A second spectrum, taken five days later, shows SY Vol brighter, probably on the decline from outburst.  This spectrum is very similar to the one presented by \cite{CieslinskiSteinerJablonski98}, showing a blue continuum and Balmer lines with broad absorption wings.  He\,{\scriptsize I}\,$\lambda$5876 emission is weakly detected in both spectra.

\begin{figure}
 \centering
 \includegraphics[width=8.8cm]{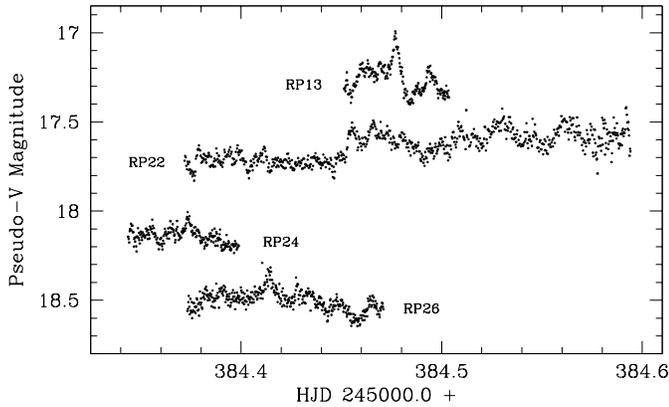}
 \caption{The light curves of SY Vol.  The run RP13, RP24, and RP26 observations are displaced vertically by -0.35, +0.3, and +0.6~mag respectively, and horizontal shifts of $-2$~d, $-3$~d, and $-4$~d respectively were applied to the RP22, RP24, and RP26 light curves.}
 \label{fig:syvollc}
\end{figure}

We took high-speed photometry of SY Vol on four occasions; these observations show the rapid flickering commonly seen in CVs.  The systems also displayed a small systematic decline in brightness over the 4 nights for which we have photometry.  This data set reveals no clear sign of an orbital modulation.  We also searched all our light curves for periodic or quasi-periodic variations on shorter time scales, but found none.  The light curves are displayed in Fig.~\ref{fig:syvollc}, although they add little to the study of \cite{surveyV}.  The low quality of the second half of run RP22 resulted from poor seeing.

The fact that all published studies of SY Vol include observations of high and low photometric states suggests that this system has a high DN outburst duty cycle.

\subsection{1RXS J092737.4-191529}
\begin{figure}
 \centering
 \includegraphics[width=8.8cm]{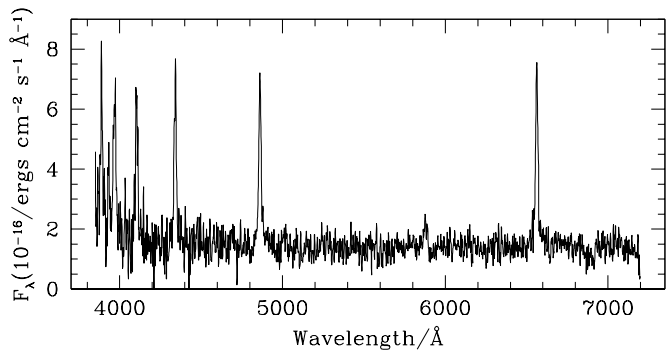}
 \caption{The average of the two spectra of 1RXS J092737.4-191529.  Broad Balmer and He\,{\scriptsize I} emission lines indicate that this object is a CV.  The strength of the emission lines and flat continuum suggest that 1RXS J092737.4-191529 is a low-$\dot{M}$ system.}
 \label{fig:0927spec}
\end{figure}

The optical counterpart of 1RXS J092737.4-191529 varied from $R=18.3$ to $R=15.4$ between the two epochs of photographic data.  The system was faint (near 18th magnitude) during all our observations.  There was no significant difference between the two spectra we took; the average of these two observations is displayed in Fig.~\ref{fig:0927spec}.  The spectrum has a flat continuum (longward of $\simeq 4\,500\,\mathrm{\AA}$) and Balmer as well as He\,{\scriptsize I}\,$\lambda$5876 emission lines.  Both the strong emission lines and the flat continuum indicate that a 1RXS J092737.4-191529 is a low-$\dot{M}$ system, probably a quiescent DN.

\begin{figure}
 \centering
 \includegraphics[width=8.8cm]{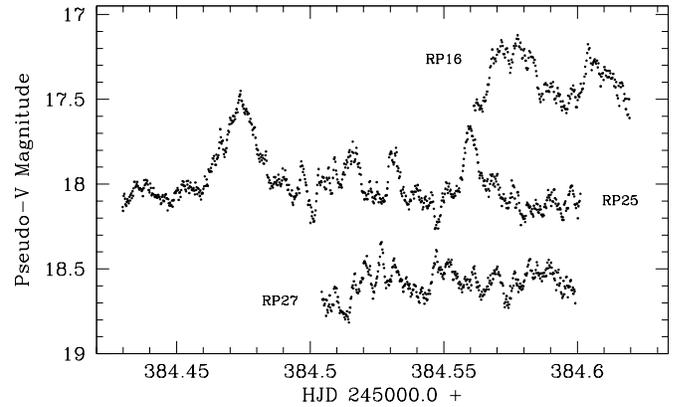}
 \caption{Our 3 longer light curves of 1RXS J092737.4-191529, showing flickering amplitudes of up to $\simeq 0.4$~mag.  The run RP27 light curve is shifted horizontally by $-4$~d and vertically by +0.5~mag.  Run RP16 and RP25 are displaced by $-0.5$~mag and $-3$~d respectively.}
 \label{fig:rxs0927lc}
\end{figure}

We obtained time-resolved photometry of 1RXS J092737.4-191529 on four nights.  The 14 January 2005 run was interrupted by cloud after less than an hour; the remaining three light curves are displayed in Fig.~\ref{fig:rxs0927lc}.  The system flickers over a range of roughly 0.8~mag.  Flickering is the hallmark of mass transfer, and its presence in the light curves confirms that 1RXS J092737.4-191529 is a CV.  The longest observation spanned more than 4~h, but reveals nothing that is easily recognised as an orbital modulation.  A Fourier analysis failed to reveal any coherent or quasi-coherent modulations in the photometry.

\section{Discussion and Summary}
\label{sec:discussion}
Requiring ROSAT detection as well as at least 2~mag variability, we selected three objects for follow-up observations.  The optical selection is based on a very incomplete sample, and we have used only two epochs of optical observations.  Two of our three targets are CVs, suggesting that this selection may be very efficient for finding CVs.  Clearly, selection based on variability will be improved by higher quality optical data and/or more epochs.  With large CCD mosaics becoming more common, wide-angle, multi-epoch surveys with much better photometric accuracy will soon be available.  Relatively deep variability surveys currently in progress are aimed at short time scale variability, and cover smaller areas (e.g. \citealt{GrootVreeswijkHuber03}; \citealt{RamsayHakala05}).  However, it will in future be possible to do all-sky surveys with $\sim 1$~d sampling \citep{Walker03}.  An obvious advantage that existing photographic archives will always have is their observational baseline of several decades.

In summary, we have presented observations of two CVs that were selected for large amplitude optical variability and X-ray emission.  1RXS J092737.4-191529 is a newly discovered system, while the optical counterpart of 1RXS J085325.4-711255 is the known DN, SY Vol.  1RXS J092737.4-191529 has the observational appearance of a low-$\dot{M}$ system, and is probably also a DN.  This is the class of CVs that our selection is the most sensitive to.

\begin{acknowledgements}
MLP acknowledges financial support from the South African National Research Foundation and the University of Southampton.  
\end{acknowledgements}


\end{document}